\documentclass[a4paper,epsfig,11pt]{article}
\usepackage{jcappub} % for details on the use of the package, please
\usepackage[T1]{fontenc} % if needed

\usepackage{graphicx,amssymb,amsmath,amsthm,amsfonts,epsfig,times,bm}
\usepackage{aas_macros}
\usepackage{cleveref}
\usepackage{amsmath,amssymb}
\usepackage{tensor}

\def\be{\begin{equation}}
\def\ee{\end{equation}}

\newcommand{\bea}{\begin{eqnarray}}
\newcommand{\eea}{\end{eqnarray}}

\title{\boldmath Reconstruction of interaction rate in Holographic dark energy}

\author[]{Ankan Mukherjee \footnote{E-mail: ankan\_ju@iiserkol.ac.in}} 

\affiliation[]{Department of Physical Sciences,~~\\Indian Institute of Science Education and Research Kolkata,~~\\Mohanpur, West Bengal-741246, India}

\abstract
{
The present work is based on the holographic dark energy model with Hubble horizon as the infrared cut-off. The interaction rate between dark energy and dark matter has been reconstructed for three different parameterizations of the deceleration parameter. Observational constraints on the model parameters have been obtained by maximum likelihood analysis using the observational Hubble parameter data (OHD), type Ia supernovab data (SNe), baryon acoustic oscillation data (BAO) and the distance prior of cosmic microwave background (CMB) namely the CMB shift parameter data (CMBShift). The interaction rate obtained in the present work remains always positive and increases with expansion. It is very similar to the result obtained by Sen and Pavon \cite{senpavon} where the interaction rate has been reconstructed for a parametrization of the dark energy equation of state. Tighter constraints on the interaction rate have been obtained in the present work as it is based on larger data sets. The nature of the dark energy equation of state parameter has also been studied for the present  models. Though the reconstruction is done from different parametrizations, the overall nature of the interaction rate is very similar in all the cases.  Different information criteria and the Bayesian evidence, which have been invoked in the context of model selection, show that the these models are at close proximity of each other. 
}

\begin{document}
%\date{\today}

\maketitle
\flushbottom

\tableofcontents
%%%%%%%%%%%%%%%%%%%%%%%%%%%%%%%%%%%%%%%%%%%%%%
\section{Introduction}\label{intro}
%%%%%%%%%%%%%%%%%%%%%%%%%%%%%%%%%%%%%%%%%%%%%%
The recent cosmic acceleration, discovered in late nineties \cite{riessacc,perlmutter}, is presently the most puzzling phenomenon of modern cosmology. It has put a question mark to the basic frame work of cosmology as there is no appropriate answer in the cosmological standard model regarding the genesis of  cosmic acceleration. Various observation like the Baryon Oscillation Spectroscopic Survey (BOSS) \cite{boss}, the continuation of supernova  cosmology project \cite{suzukiunion21}, the Dark Energy Survey \cite{des}, the mapping of the universe from the multi wavelength observations of the Sloan Digital Sky Survey (SDSS) \cite{sdssmap}, the observation of cosmic microwave background (CMB) from WMAP, Planck \cite{wmaplanck,planck15} etc. are directed towards different aspects of the imprints of evolutions. The basic endeavour is to have a combination of different observations to understand the evolution history and to find the reason behind the cosmic acceleration. The unprecedented improvements in cosmological observations have upgraded the observational data to a higher level of precision and much tighter constraints on various cosmological models have been achieved. But still now, there is hardly any signature to identify the actual reason of cosmic acceleration.

\par In literature, there are various prescriptions to explain this phenomenon. These can be classified into two classes. One is the {\it dark energy}, an exotic component introduced in the energy budget of the universe, which can generate the cosmic acceleration with its characteristic negative pressure. For the dark energy models, the General Relativity (GR) is taken as the proper theory of gravity. The other way to look for the solution through the modification of GR such as $f(R)$ gravity models \cite{Cappoz,cardu,voldn,nojod03,car05,menaosan,nojod06,dasban,nojod09}, scalar-tensor theory \cite{bertmar,banpav1,banpav2,sensen,motabar1,motabar2,sndasban}, different higher dimensional gravity theories \cite{dedv,deff,nojodsami,dagapo,hossamw,bambak} etc.

\par In the context of dark energy, the simplest and consistent with most of the observations is the cosmological constant model where the constant vacuum energy density serves as the dark energy candidate. But there are different issues related to the cosmological constant model. There is a huge discrepancy between the observationally estimated value of cosmological constant and the value calculated from quantum field theory. It also suffers from the cosmic coincidence problem. Comprehensive discussions on the cosmological constant model  are there by Carroll \cite{carrollLCDM} and by Padmanabhan \cite{padmanabhanLCDM} where different issues have been emphasised in great details. Due to these issues related to the cosmological constant, time varying dark energy models also warrant attention. Review articles with comprehensive discussion on different dark energy models are there in literature \cite{sahnist,peerat,copsa,marj}. The present trend in cosmological modelling is {\it reconstruction} which is a reverse way of finding the viable model of cosmic evolution. The idea is to adopt a viable evolution scenario and then to find the behaviour of the relevant cosmological quantities and to estimate the values of the parameters associated to the model. Reconstruction of dark energy models has earlier been discussed by Starobinsky \cite{staraa}, Huterer and Turner \cite{hutur1,hutur2} and by Saini et al. \cite{saisr}. With the recent unprecedented improvement in the cosmological observations, the dark energy models are becoming highly constrained. Parametric and non-parametric, both types of reconstructions with various updated observational data are giving more and more precise estimation of the dark energy parameters \cite{xiali,hazd,hols,despop,mukherjeeweff}. Reconstruction of kinematical quantities like the deceleration parameter, cosmological jerk parameter have been discussed in reference \cite{rapK,luongojerk,zhaijerk,amnbjerk}. 

\par In most of the dark energy models, the dark matter and dark energy are allowed to have independent conservation, ignoring the possibility of interaction between them. Das {\it et al.} \cite{dasint} and Amendola {\it et al.} \cite{amenint} have shown that the the prior ignorance of the interaction between the dark energy and dark matter might cause some misleading results. It has been argued that the phantom nature of dark energy might be consequence of ignoring the possibility of interaction between dark energy and dark matter \cite{dasint,amenint}. There are also good number of investigations in the literature on the interacting dark energy models \cite{ame1,ame2,billco,zimpav,ameq,herpav}. Interaction between Brans-Dicke scalar field and quintessence has been discussed by Banerjee and Das \cite{bandasint}. Some recent attempts to find the constraint on interacting dark energy models from recent observational data are by Paliathanasis and Tsamparlis \cite{paliats}, by Pan {\it et al.} \cite{panint}, Nunes {\it et al.} \cite{nunespanint} and by Murgia {\it et al.} \cite{murgiagariazzo}. In a recent analysis by Salvatelli {\it et al.} \cite{salvatellisaid}, it has been shown that an interacting vacuum cosmology, where the coupling strength varies with redshift, can be a possible solution to the tension in $\Lambda$CDM model between the CMB data and the measurement of linear growth rate of large scale structure (LSS) from redshift-space distortion (RSD) data.    Non-parametric reconstruction of interacting dark energy has been discussed by Yang {\it et al.} \cite{yanguochi}. Recent review on dark matter dark energy interaction by Wang {\it et al.} \cite{wangabdallapav} presents a comprehensive discussion on different aspects and theoretical challenges of interacting dark energy. 

\par The present work is the reconstruction of the interaction rate of {\it holographic dark energy}. The basic idea of holographic dark energy is based on fundamental thermodynamic consideration, namely the {\it holographic principal},  introduced by 't Hooft \cite{hooft} and Susskind \cite{susskind}. To avoid the violation of the second law of thermodynamics in the context of quantum theory of gravity, Bekenstein suggested that
the maximum entropy of the system should be proportional to its area instead  of its volume \cite{bekenstein}. Form this idea, t'Hooft conjectured that the phenomena within a volume can be explained by the set of degrees of freedom residing on its boundary and the degrees of freedom of a system is determined by the area of the boundary instead of the volume of the system. In quantum field theory it relates a short distance cut-off (ultraviolet (UV) cut-off) to a long distance cut off (infra red (IR) cut-off) in the limit set by the formation of a black hole \cite{cohenhol}. The total quantum zero point energy of a system should not exceed the mass of a black hole of the same size. If $\rho_{\Lambda}$ be the quantum zero point energy density caused by the short distance cut-off, the total energy is $L^3\rho_{\Lambda}$, where $L$ is the size  of the system. Thus it can be written as \cite{limhol},

\be
L^3\rho_{\Lambda}\leq LM_P^2, 
\ee 
where $M_P^2=(8\pi G)^{-1}$. The inequality saturates for the largest allowed value of the system size $L$, which is the long distance cut-off or the infrared cut-off. Thus the energy density $\rho_{\Lambda}$ be proportional to inverse square of the infra red cut-off. This idea have been adopted in the context of dark energy by Li \cite{limhol}. Thus the holographic dark energy density is written as,
\be
\rho_{H}=3C^2M_P^2/L^2,
\label{rhoh}
\ee 
where $C^2$ is a dimensionless constant. Different attempts are  there in literature with different selections of the infrared cut-off length scale, the particle horizon \cite{parhorizon,parhorizon2}, the future event horizon \cite{limhol,futhorizon1,futhorizon2,futhorizon3,futhorizon4,futhorizon5,futhorizon6} and the Hubble horizon \cite{hubhorizon} etc. Holographic dark energy in Brans-Dicke theory has been discussed by Banerjee and Pavon \cite{banpavholo}. Xu has studied holographic dark energy with the Hubble horizon  cut-off with constant as well as time varying coupling parameter ($C^2$) \cite{xuholo}. A comparative study of the holographic dark energy  with different length scale cut-off has been carried out by Campo {\it et al.} \cite{campoholo}. Recently Hu {\it et al.} \cite{huyholo} has attempted to built up the model combining cosmological constant and holographic energy density. Holographic dark energy from minimal supergravity has been discussed by Landim \cite{landimholo}. Stability analysis of holographic dark energy model has been discussed by Banerjee and Roy \cite{banroyholo}. 

\par In the present work, the Hubble horizon has been adopted as the infrared cut-off for the holographic dark energy meaning the cutoff length scale $L=(H)^{-1}$, where $H$ is the Hubble parameter. Now it is imperative to note that the holographic dark energy models with Hubble horizon cut off can generate late time acceleration along with the matter dominated decelerated expansion phase in the past only if there is some interaction between dark energy and dark mater.

\par In the present work, the interaction rate of holographic dark energy has been reconstructed from three different parameterizations of the deceleration parameter. The expressions of Hubble parameter obtained for these models hardly give any indication towards the independent conservation of dark matter and dark energy. The prime endeavour of the present work is to study the nature of interaction and the evolution of the interaction rate for these three models assuming the holographic dark energy with Hubble horizon as the IR cut-off. Reconstruction of interaction rate in holographic dark energy has earlier been discussed by Sen and Pavon \cite{senpavon}, where the interaction rate has been reconstructed assuming a particular form of the dark energy equation of state.   

\par In section \ref{reconst}, the reconstruction of the interaction rate for these three models have been discussed. In section \ref{obsdata}, a brief discussion about the observational data sets, used in the statistical analysis, have been presented. Section \ref{results} presents the results of statistical analysis of the models including the constraints on the model parameters and also the constraints on the evolution of holographic interaction rate. In section \ref{bayesian}, a Bayesian analysis has been presented to select the preferred model among these three, discussed in the present work.  Finally, in section \ref{discussion}, an overall discussion about the results obtained has been presented.

\section{Reconstruction of the interaction rate}
\label{reconst}
The metric of a homogeneous and isotropic universe with a spatially flat geometry is written as,
\be
ds^2=-dt^2+a^2(t)[dr^2+r^2d\Omega^2],
\ee
where $a(t)$, the time dependent coefficient of the spatial part of the metric, is  called the {\it scale factor}. Now the Hubble parameter is defined as $H=\frac{\dot{a}}{a}$, where the dot denotes the derivative with respect to time. The Friedmann equations, written in terms of $H$, are
\be
3H^2=8\pi G(\rho_m+\rho_{DE}),
\label{friedmann1}
\ee
and
\be
2\dot{H}+3H^2=-8\pi G(p_{DE}),
\label{friedmann2}
\ee
where $\rho_m$ is  the energy density of the pressureless dust matter and $\rho_{DE}$ and $p_{DE}$ are respectively the energy density and pressure of the dark energy. Now from contracted Bianchi identity, the conservation equation can be wriieten as,
\be
\dot{\rho}_{total}+3H(\rho_{total}+p_{total})=0,
\label{totconservation}
\ee
where $\rho_{total}=\rho_m+\rho_{DE}$ and $p_{total}=p_{DE}$ as the dark matter is pressureless. Now the conservation equation (equation \ref{totconservation}) can be decomposed into two parts,

\be
\dot{\rho}_m+3H\rho_m=Q,
\label{matterconservation}
\ee
and 
\be
\dot{\rho}_{DE}+3H(1+w_{DE})\rho_{DE}=-Q,
\label{darkenergyconservation}
\ee
where $w_{DE}$ is the equation of state parameter  of dark energy and the $Q$ is the interaction term. If there is no interaction between dark energy and dark matter, then the interaction term $Q=0$, and the matter evolves as, $\rho_m\propto\frac{1}{a^3}$.

\par The prime goal of the present work is to study the interaction assuming a holographic dark energy with Hubble horizon as the IR cut off. Holographic dark energy models with the Hubble horizon $H^{-1}$ as the IR cut-off require the interaction between the dark energy and dark matter to generate the late time acceleration. The dark energy density $\rho_{DE}$ for a holographic model with the Hubble horizon as the IR cut-off (denoted as $\rho_H$) is given, according to equation (\ref{rhoh}) as,
\be 
\rho_H=3C^2M_P^2H^2,
\ee
where $C$, the coupling parameter is assumed to be a constant in the present work and $M_P=\frac{1}{\sqrt{8\pi G}}$. Now the interaction term $Q$ is written as, $Q=\rho_H\Gamma$, where $\Gamma$ is the rate at which the energy exchange occurs between dark energy and dark matter. The ratio of dark mater and dark energy density, sometimes called the {\it coincidence parameter}, is written as, $r=\rho_m/\rho_H$, and its time derivative can be expressed as \cite{senpavon},
\be
\dot{r}=(1+r)\Big[3Hw_{DE}\frac{r}{1+r}+\Gamma\Big].
\label{rdot}
\ee 
For a spatially flat geometry, it can also be shown that the ration $r$ remains constant for a holographic dark energy with Hubble horizon as the  IR cut-off. As the ratio of dark matter and dark energy remains constant in this case, it can potentially convey the answer to the cosmic coincidence problem. But it might be confusing as one may think that it contradicts the standard scenario of structure formation during the dark matter dominated epoch. Actually this is not the case. The matter dominated phase is automatically recovered as the interaction rate is very small at high and moderate redshift and thus the dark energy equation of state resembles the non-relativistic matter \cite{hubhorizon,hubhorizon2}. For a constant value of $r$, $\dot{r}=0$, from which the interaction rate can be expressed using equation (\ref{rdot}) as,
\be
\Gamma=-3Hr\frac{w_{DE}}{1+r}.
\label{intrate}
\ee  
The effective or total equation of state parameter ($w_{eff}=\frac{p_{total}}{\rho_{total}}$), is related to the dark energy equation of state parameter as,
\be 
w_{eff}=\frac{w_{DE}}{1+r}.
\label{weff}
\ee
Finally the interaction can be written as, 

\be
\Gamma=-3Hrw_{eff},
\ee 
and representing it in a dimensionless way,
\be
\frac{\Gamma}{3H_0}=-(H/H_0)rw_{eff}.
\label{intratere}
\ee
In the present work, the interaction rate has been reconstructed for three different parameterizations of the deceleratin parameter. The expression of the Hubble parameter obtained for these models hardly give any indication of the independent evolution of dark energy and dark matter, thus these parameterizations are useful to study the interaction. These parameterizations of deceleration parameter have been discussed in the following. It worth mentioning that for the reconstruction of the interaction rate, it is required to fix the value of the coincident parameter $r$. The value of $r$ is taken according to the recent measurement of the dark energy density parameter $\Omega_{DE0}$ from Planck using Planck+WP+highL+BAO \cite{planck} as for spatially flat universe $r$ can be written as $r=(1-\Omega_{DE0})/\Omega_{DE0}$. It is imperative to note that the interaction rate $\Gamma$ does not depends upon the coupling parameter ($C^2$). The effective equation of state parameter ($w_{eff}(z)$) can be obtained from the Hubble parameter using the Friedmann equations (equation (\ref{friedmann1}) and (\ref{friedmann2})).

\par The deceleration parameter, a dimensionless representation of the second order time derivative of  the scale factor, is defined as $q=-\frac{1}{H^2}\frac{\ddot{a}}{a}$. It can also be written using redshift $z$ as the argument of differentiation as,
\be
q(z)=-1+\frac{1}{2}(1+z)\frac{(H^2)'}{H^2}.
\ee 
The parametric forms of the deceleration parameter, adopted in the present work, are given as,
\be
Model ~~I.~~~~~~~~q(z)=q_1+\frac{q_2}{(1+z)^2},
\ee
\be
Model ~~II.~~~~~~~~q(z)=\frac{1}{2}+\frac{q_1+q_2z}{(1+z)^2},
\ee
\be
Model ~~III.~~~~~~~~q(z)=-1+\frac{q_1(1+z)^2}{q_2+(1+z)^2},
\ee
where $q_1$ and $q_2$ are the parameters for the models. However, $q_1$ and $q_2$ do not have the same physical significance in the three different models. The second model of deceleration parameter adopted in the present work has already been discussed by Gong and Wang \cite{gongwang,gongwang2} in the context of reconstruction of the late time dynamics of the Universe. The parametrization of Model III has some similarity with one of the parametrizations based on thermodynamic requirement discussed by Campo {\it et al.} \cite{campopavon}. The variation of deceleration parameter at low redshift is higher for Model I and Model II than Model III. Thus parametrization of Model III is significatly different from other two. The expressions of Hubble parameter scaled by its present value for the models yield to be
\be
Model~~I.  ~~~~~~h^2(z)=\frac{H^2(z)}{H_0^2}=(1+z)^{2(1+q_1)}\exp{\Bigg[-q_2\Bigg(\frac{1}{(1+z)^2}-1\Bigg)\Bigg]},
\label{hubbleparameter1}
\ee
\be
Model ~~ II. ~~~h^2(z)=\frac{H^2(z)}{H_0^2}=(1+z)^3\exp{\Bigg[\frac{q_2-q_1}{(1+z)^2}-\frac{2q_2}{(1+z)}+(q_1+q_2)\Bigg]},
\label{hubbleparameter2}
\ee
\be
Model ~~ III. ~~~h^2(z)=\frac{H^2(z)}{H_0^2}=\left(\frac{q_2+(1+z)^2}{1+q_2}\right)^{q_1},
\label{hubbleparameter3}
\ee
and consequently the effective equation of state parameter ($w_{eff}(z)$) for the models are expressed as
\be
Model~~I. ~~~~ w_{eff}(z)=-1+\frac{2}{3}\Bigg[(1+q_1)+\frac{q_2}{(1+z)^2}\Bigg],
\ee
\be
Model ~~ II. ~~~w_{eff}(z)=-1+\frac{1}{3}\Bigg[3+\frac{2q_2}{(1+z)}-\frac{2(q_2-q_1)}{(1+z)^2}\Bigg],
\ee
\be
Model ~~ III. ~~~w_{eff}(z)=-1+\frac{2}{3}\left(\frac{q_1(1+z)^2}{q_2+(1+z)^2}\right). 
\ee

Utilizing the expression of the effective equation of state, the interaction rate of holographic dark energy can be reconstructed using equation (\ref{intratere}). It is also worth mentioning in the context is that this expressions of Hubble parameter (equation (\ref{hubbleparameter1}),(\ref{hubbleparameter2}) and (\ref{hubbleparameter3})) hardly give any indication regarding the independent conservation of dark matter and dark energy as the dark matter and dark energy components are not separately identified in the expressions of $h^2(z)$ in equations (\ref{hubbleparameter1}), (\ref{hubbleparameter2}) and (\ref{hubbleparameter3}).

\section{Observational data}
\label{obsdata}
Different observational data sets have been utilized for the statistical analysis of the models in the present work. These are the observational Hubble data (OHD), distance modulus data from type Ia supernove (SNe), baryon acoustic oscillation (BAO) data along with the value of acoustic scale at photon electron decoupling and the ratio of comoving sound horizon at decoupling and at drag epoch estimated from Cosmic Microwave Background 
(CMB) radiation power spectrum and the CMB distance prior namely the CMB shift parameter (CMBShift) data. The data sets are briefly discussed in the following. The discussion about the observational data has also been presented in a very similar fashion in \cite{mukherjeeweff,amnbjerk}

\subsection{Observational Hubble parameter data}
The data of Hubble parameter measurement by different groups have been used in the present analysis. Hubble parameter $H(z)$ can be estimaed from the measurement of differential of redshift $z$ with  respect to cosmic time $t$ as
\be
H(z)=-\frac{1}{(1+z)}\frac{dz}{dt}.
\label{Hz}
\ee
The differential age of galaxies have been used as an estimator of $dz/dt$ by Simon {\it et al.}  \cite{Simon2005}. Measurement of cosmic expansion history using red-enveloped galaxies was done by Stern {\it et al} \cite{Stern2010} and by Chuang and Wang \cite{ChuangWang2013}. Measurement of expansion history from WiggleZ Dark Energy Survey has been discussed by Blake {\it et al.} \cite{Blake2012}. Measurement of Hubble parameter at low redshift  using the differential age method along with Sloan Digital Sky Survey (SDSS) data have been presented by Zhang {\it et al}  \cite{Zhang2014}. Compilation of observational Hubble parameter measurement has been presented by Moresco {\it et al} \cite{Moresco2012}. Finally, the measurement of Hubble parameter at $z=2.34$ by Delubac {\it et al} \cite{Delubac2015} has also been used in the present analysis. The measurement of $H_0$ from Planck+WP+highL+BAO \cite{planck} has also been used in the analysis. The  relevant $\chi^2$ is defined as
\be
\chi^2_{{\tiny OHD}}=\sum_{i}\frac{[H_{obs}(z_i)-H_{th}(z_i,\{\theta\})]^2}{\sigma_i^2},
\label{chiOHD}
\ee 
where $H_{obs}$ is the observed value of the Hubble parameter, $H_{th}$ is theoretical one and $\sigma_i$ is the uncertainty  associated to the $i^{th}$ measurement. The $\chi^2$ is a function of the set of  model parameters $\{\theta\}$.

\subsection{Type Ia supernova data} 
The measurement of the distance modulus of type Ia supernova of the most widely used data set in the modelling of late time cosmic acceleration. The distance modulus of type Ia supernova is the difference between the apparent magnitude ($m_B$) and absolute magnitude ($M_B$) of the B-band of the observed spectrum. It is defined as
\be
\mu(z)=5\log_{10}{\Bigg(\frac{d_L(z)}{1Mpc}\Bigg)}+25,
\ee
where the $d_L(z)$ is the luminosity distance and in a spatially flat FRW universe it is defined as
\be
d_L(z)=(1+z)\int_0^z\frac{dz'}{H(z')}.
\ee
In the present work, the 31 binned data sample of the recent joint lightcurve analysis (jla) \cite{jla} has been utilized. To account for the correlation between different bins, The formalism discussed by Farooq, Mania and Ratra \cite{farooqmrat} to account for the correlation between different redshift bins, has been adopted. The $\chi_{SNe}^2$ has been defined as
\be
\chi_{SNe}^2=A(\{\theta\})-\frac{B^2(\{\theta\})}{C}-\frac{2\ln{10}}{5C}B(\{\theta\})-Q,
\label{chiSNe}
\ee     
where 
\be 
A(\{\theta\})=\sum_{\alpha,\beta}(\mu_{th}-\mu_{obs})_{\alpha}(Cov)^{-1}_{\alpha\beta}(\mu_{th}-\mu_{obs})_{\beta},
\ee
\be
B(\{\theta\})=\sum_{\alpha}(\mu_{th}-\mu_{obs})_{\alpha}\sum_{\beta}(Cov)^{-1}_{\alpha\beta},
\ee
\be
C=\sum_{\alpha,\beta}(Cov)^{-1}_{\alpha\beta},
\ee
and the $Cov$ is the $31\times31$ covarience matrix of the binning. The constant term $Q$ can be ignored during the analysis as it is independent of the model parameters.

\subsection{Baryon acoustic oscillation data}
The baryon acoustic oscillation (BAO) data have been used in the present analysis in combination with the Planck \cite{planck,wangwangCMB} measurement of the  {\it acoustic scale ($l_A$)}, the {\it comoving sound horizon ($r_s$)} at photon decoupling epoch ($z_*$) and at drag epoch ($z_d$). The BAO data have been used in the form of a ratio of the {\it comoving angular diameter distance} at decoupling ($d_A(z_*)=c\int_0^{z_*} \frac{dz'}{H(z')}$),  and the {\it dilation scale} ($D_V(z)=[czd_A^2(z)/H(z)]^{\frac{1}{3}}$). Three mutually uncorrelated measurements of $\frac{r_s(z_d)}{D_V(z)}$ ( 6dF Galax Survey at redshift $z=0.106$ \cite{beutlerbao}, Baryon Oscillation Spectroscopic Survey (BOSS) at redshift $z=0.32$ (BOSS LOWZ) and at redshift $z=0.57$ (BOSS CMASS) \cite{andersonbao}) have been adopted in the present analysis.  Table \ref{tableBAO} contains the values of $\Big(\frac{r_s(z_d)}{D_V(z_{BAO})}\Big)$ and finally the $\Big(\frac{d_A(z_*)}{D_V(z_{BAO})}\Big)$ at three different redshift of BAO measurement.

%%%%%%%%%%%%%%%%%%%%%%%%%%%%%%%%%
\begin{table*}
\caption{\small BAO/CMB data table}.
\begin{center}
\resizebox{0.6\textwidth}{!}{  
\begin{tabular}{ |c |c |c |c |} 
 \hline
  \hline
 $z_{BAO}$ & 0.106 & 0.32 & 0.57 \\ 
 \hline
 \hline
 $\frac{r_s(z_d)}{D_V(z_{BAO})}$ & 0.3228$\pm$0.0205 & 0.1167$\pm$0.0028 & 0.0718$\pm$0.0010 \\ 
 \hline
 $\frac{d_A(z_*)}{D_V(z_{BAO})}\frac{r_s(z_d)}{r_s(z_{*})}$ & 31.01$\pm$1.99 & 11.21$\pm$0.28  & 6.90$\pm$0.10 \\ 
 \hline
 $\frac{d_A(z_*)}{D_V(z_{BAO})}$ & 30.43$\pm$2.22 & 11.00$\pm$0.37  & 6.77$\pm$0.16\\ 
 \hline
\hline

\end{tabular}
}
\end{center}
\label{tableBAO}
\end{table*}
%%%%%%%%%%%%%%%%%%%%%%%%%%%%%%%%%%%%%%%

The relevant $\chi^2$, namely $\chi^2_{BAO}$, is defined as:
\be
\chi^2_{BAO}={\bf X^{t}C^{-1}X},
\label{chibao}
\ee
where
\[
{\bf X}=
\left( {\begin{array}{c}
   \frac{d_A(z_*)}{D_V(0.106)}-30.43   \\
   \frac{d_A(z_*)}{D_V(0.2)}-11.00     \\
   \frac{d_A(z_*)}{D_V(0.35)}-6.77  \\
\end{array} } \right)
\]
and ${\bf C^{-1}}$ is the inverse of the covariance matrix. As the three measurements are mutually uncorrelated, the covariance matrix is diagonal.

\subsection{CMB shift parameter data:}
The CMB shift parameter, which is related to the position of the first acoustic peak in power spectrum of the temperature anisotropy of the Cosmic Microwave Background (CMB) radiation, is defined in a spatially flat uiverse as,
\be
{\mathcal R}=\sqrt{\Omega_{m0}}\int_0^{z_*}\frac{dz}{h(z)},
\label{cmbshift}
\ee  where $\Omega_{m0}$ is the matter density parameter, $z_*$ is the redshift at photon decoupling and $h(z)=\frac{H(z)}{H_0}$ (where $H_0$ be the present value of Hubble parameter). In general it is efficient to ensure tighter constraints on the model parameters if used in combination with other observational data. The value of CMB shift parameter is not directly measured from CMB observation. The value is estimated from the CMB data along with some fiducial assumption about  the background cosmology. The $\chi^2_{\tiny CMBShift}$ is defined as 
\be 
\chi^2_{\tiny CMBShift}=\frac{({\mathcal R}_{obs}-{\mathcal R}_{th}(z_*))^2}{\sigma^2},
\ee
where ${\mathcal R}_{obs}$ is the value of the CMB shift parameter, estimated from observation and $\sigma$ is the corresponding uncertainty. In this work, the value of CMB shift parameter estimated from Planck data \cite{wangwangCMB} has been used. It is imperative to mention that in the present analysis the value  $\Omega_{m0}$ is taken according to recent estimation from Planck \cite{planck}.

\section{Results of statistical analysis}
\label{results}

An indispensable part of reconstruction is the estimation of the values of the model parameters from observational data. The values of the model parameters have been estimated by $\chi^2$ minimization. Normally the $\chi^2$ is defined as 
\be 
\chi^2(\{\theta\})=\sum_i\frac{[\epsilon_{obs}(z_i)-\epsilon_{th}(z_i,\{\theta\})]^2}{\sigma_i^2},
\ee 
where $\epsilon_{obs}$ is the value of the observable measured at redshift $z_i$, $\epsilon_{th}$ from of the observable quantity as a function of the set of model parameters $\{\theta\}$ and $\sigma_i$ is the uncertainty associated to the measurement at $z_i$. In case of supernova distance modulus data and BAO data, the relevant $\chi^2$ are defined in a complicated way to incorporate the associated correlation matrix (equations (\ref{chiSNe}) and (\ref{chibao})). The combined analysis has been carried out by adding the $\chi^2$ of the individual data sets taken into account for that particular combination. The combined $\chi^2$ is defined as,
\be
\chi^2_{combined}=\sum_d\chi^2_d,
\ee 
where $d$ represents the individual data set. The $\chi^2$ associated to different data sets have been discussed in section \ref{obsdata}.

The $\chi^2$ minimization, which is equivalent to the maximum likelihood analysis, has been adopted in  the present work for the estimation of the parameter values. The likelihood is defined as, 

\be
{\mathcal{L}}(\{\theta\})=\exp{\Big(-\frac{\chi^2}{2}\Big)}.
\ee

%%%%%%%%%%%%%%%%%%%%%
\begin{figure}[tb]
\begin{center}
\includegraphics[angle=0, width=0.3\textwidth]{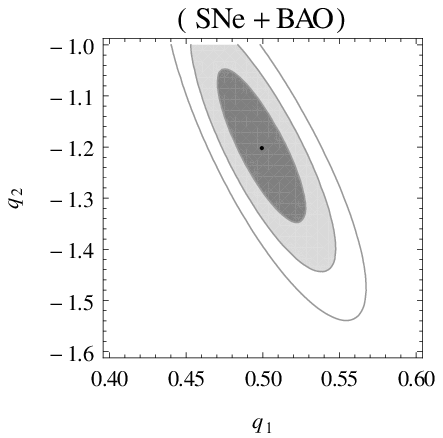}
\includegraphics[angle=0, width=0.3\textwidth]{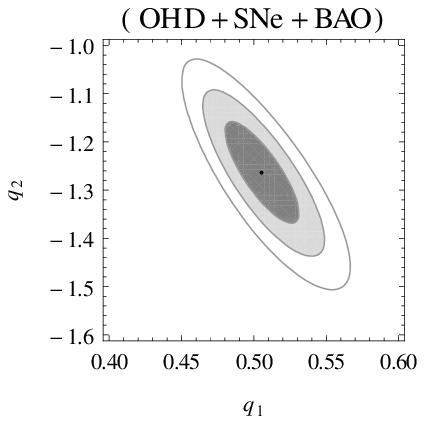}
\includegraphics[angle=0, width=0.3\textwidth]{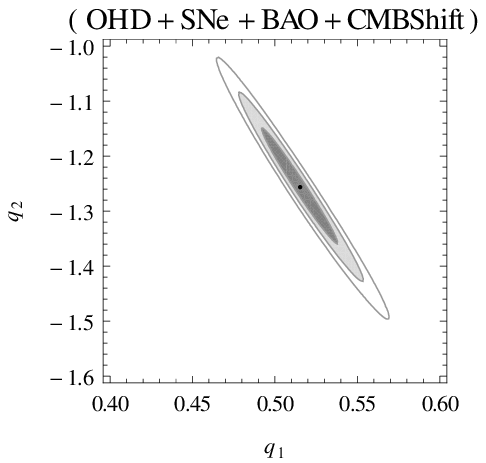}
\end{center}
\caption{{\small The confidence contours on the 2D parameter space of Model I. The 1$\sigma$, 2$\sigma$, and 3$\sigma$ confidence contours are presented from inner to outer regions, and the central black dots represent the corresponding best fit points. The left panel is obtained for SNe+BAO, the moddle panel is obtained for OHD+SNe+BAO and  the right panel is for OHD+SNe+BAO+CMBShift.}}
\label{contourplotqz1}
\end{figure}
%%%%%%%%%%%%%%%%%%%%%%
%%%%%%%%%%%%%%%%%%%%%
\begin{figure}[tb]
\begin{center}
\includegraphics[angle=0, width=0.32\textwidth]{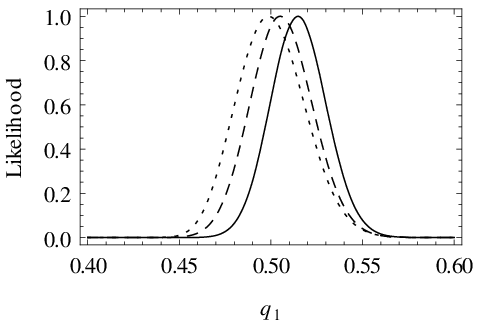}
\includegraphics[angle=0, width=0.32\textwidth]{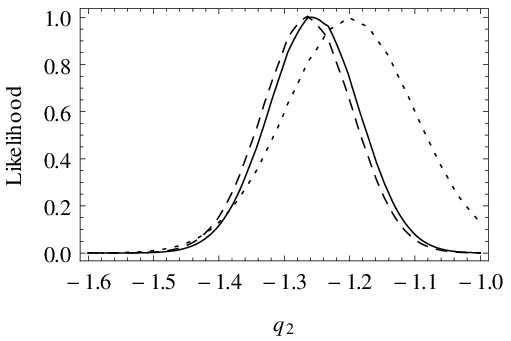}
\end{center}
\caption{{\small The marginalized likelihood as function of the model parameters $q_1$  (left panel) and $q_2$ (right panel) for Model I. The dotted curves are obtained for SNe+BAO, the dashed curves are obtained for OHD+SNe+BAO and the solid curves are obtained for OHD+SNe+BA+CMBShift.}}
\label{likelihoodplotqz1}
\end{figure}
%%%%%%%%%%%%%%%%%%%%%%

%%%%%%%%%%%%%%%%%%%%%
\begin{figure}[tb]
\begin{center}
\includegraphics[angle=0, width=0.3\textwidth]{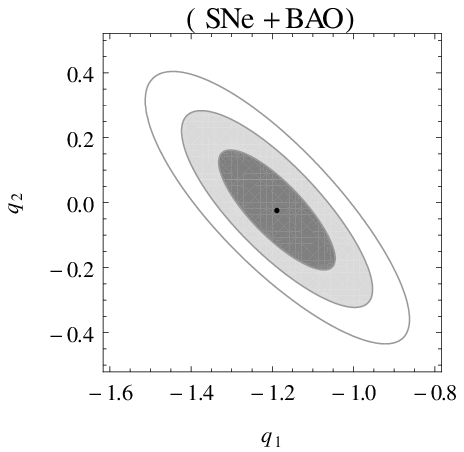}
\includegraphics[angle=0, width=0.3\textwidth]{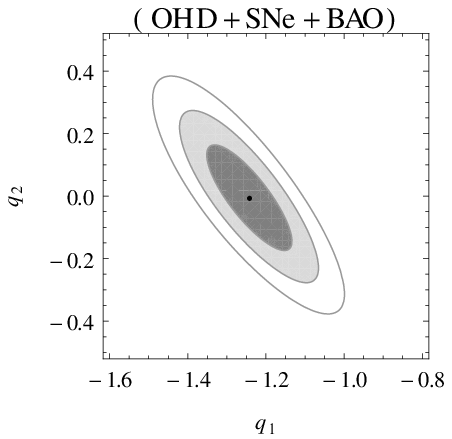}
\includegraphics[angle=0, width=0.3\textwidth]{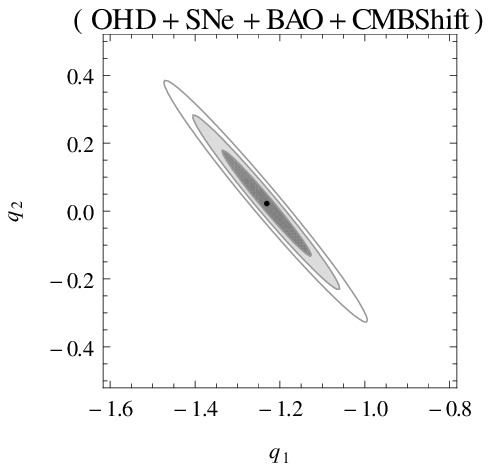}
\end{center}
\caption{{\small The confidence contours on the 2D parameter space of Model II. The 1$\sigma$, 2$\sigma$, and 3$\sigma$ confidence contours are presented from inner to outer regions, and the central black dots represent the corresponding best fit points. The left panel is obtained for SNe+BAO, the middle panel is obtained for OHD+SNe+BAO and the right panel is for OHD+SNe+BAO+CMBShift.}}
\label{contourplotqz2}
\end{figure}
%%%%%%%%%%%%%%%%%%%%%%
%%%%%%%%%%%%%%%%%%%%%
\begin{figure}[tb]
\begin{center}
\includegraphics[angle=0, width=0.32\textwidth]{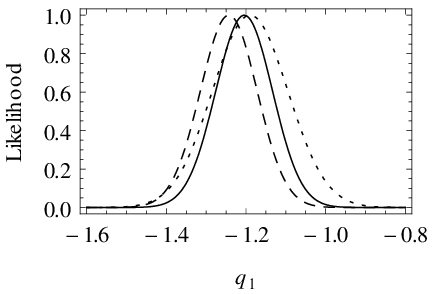}
\includegraphics[angle=0, width=0.32\textwidth]{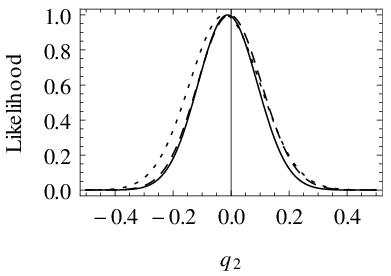}
\end{center}
\caption{{\small  The marginalized likelihood as function of the model parameters $q_1$  (left panel) and $q_2$ (right panel) for Model II. The dotted curves are obtained for SNe+BAO, the dashed curves are obtained for OHD+SNe+BAO and the solid curves are obtained for OHD+SNe+BA+CMBShift.}}
\label{likelihoodplotqz2}
\end{figure}
%%%%%%%%%%%%%%%%%%%%%%

%%%%%%%%%%%%%%%%%%%%%%%%%
\begin{table}[h!]
\caption{{\small  Results of statistical analysis of Model I with different combinations of the data sets. The value of $\chi^2_{min}/d.o.f.$ and the best fit values of the parameters along with the  associated 1$\sigma$ uncertainties are presented.}}
\begin{center}
\resizebox{0.65\textwidth}{!}{  
\begin{tabular}{ c |c |c c } 
\hline
 \hline
  Data & $\chi^2_{min}/d.o.f.$  & $q_1$ & $q_2$ \\ 
 \hline
  SNe+BAO & 35.18/28 & 0.499$\pm$0.051 & -1.202$\pm$0.367\\ 

  OHD+SNe+BAO & 50.57/54 & 0.505$\pm$0.014 & -1.264$\pm$0.064\\ 

  OHD+SNe+BAO+CMBShift & 51.97/52 & 0.515$\pm$0.013 & -1.256$\pm$0.062\\ 
 \hline
\hline
\end{tabular}
}
\end{center}

\label{tablemod1}
\end{table}
%%%%%%%%%%%%%%%%%%%%%%%%%%%%%%%%%%%

%%%%%%%%%%%%%%%%%%%%%%%%%
\begin{table}[h!]
\caption{{\small  Results of statistical analysis of Model II with different combinations of the data sets. The value of $\chi^2_{min}/d.o.f.$ and the best fit values of the parameters along with the  associated 1$\sigma$ uncertainties are presented.}}
\begin{center}
\resizebox{0.65\textwidth}{!}{  
\begin{tabular}{ c |c |c c } 
\hline
 \hline
  Data & $\chi^2_{min}/d.o.f.$  & $q_1$ & $q_2$ \\ 
 \hline
  SNe+BAO & 35.18/28 & -1.189$\pm$0.067 & -0.024$\pm$0.086\\ 

  OHD+SNe+BAO & 50.64/54 & -1.242$\pm$0.050 & -0.007$\pm$0.078\\ 

  OHD+SNe+BAO+CMBShift & 51.17/52 & -1.231$\pm$0.049 & 0.022$\pm$0.073\\ 
 \hline
\hline
\end{tabular}
}
\end{center}

\label{tablemod2}
\end{table}
%%%%%%%%%%%%%%%%%%%%%%%%%%%%%%%%%%%

%%%%%%%%%%%%%%%%%%%%%
\begin{figure}[tb]
\begin{center}
\includegraphics[angle=0, width=0.3\textwidth]{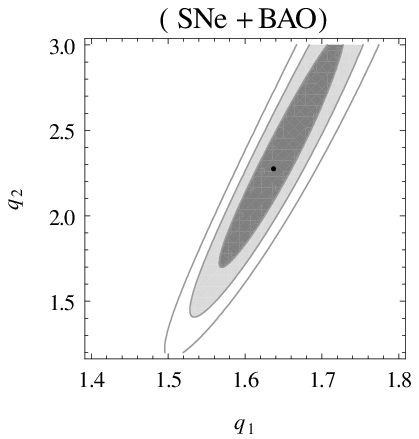}
\includegraphics[angle=0, width=0.3\textwidth]{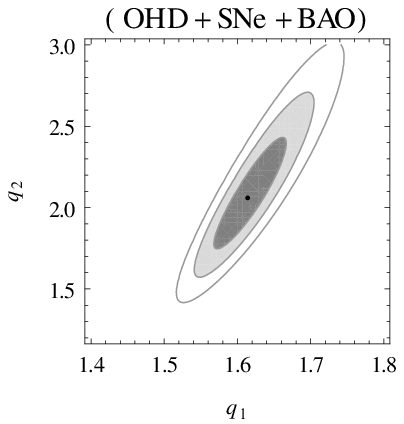}
\includegraphics[angle=0, width=0.3\textwidth]{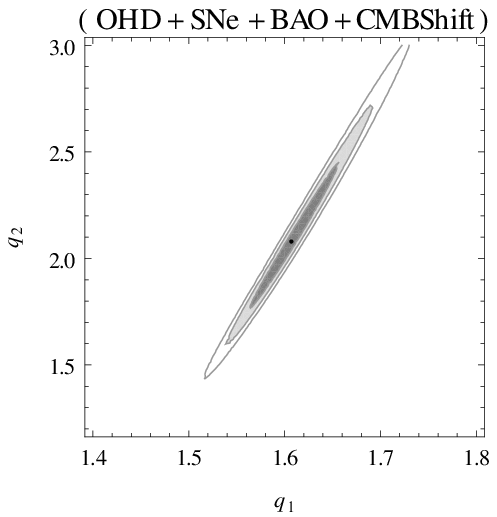}
\end{center}
\caption{{\small The confidence contours on the 2D parameter space of Model III. The 1$\sigma$, 2$\sigma$, and 3$\sigma$ confidence contours are presented from inner to outer regions, and the central black dots represent the corresponding best fit points. The left panel is obtained for SNe+BAO, the middle panel is obtained for OHD+SNe+BAO and the right panel is for OHD+SNe+BAO+CMBShift.}}
\label{contourplotqz3}
\end{figure}
%%%%%%%%%%%%%%%%%%%%%%
%%%%%%%%%%%%%%%%%%%%%
\begin{figure}[tb]
\begin{center}
\includegraphics[angle=0, width=0.32\textwidth]{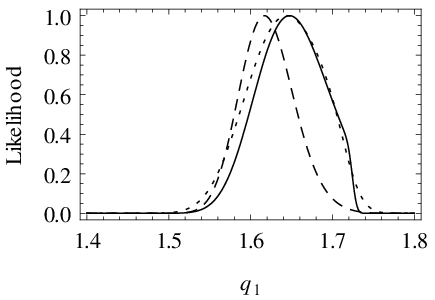}
\includegraphics[angle=0, width=0.32\textwidth]{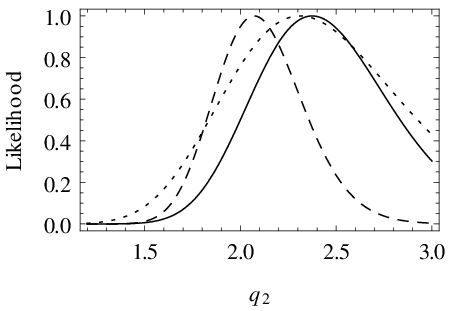}
\end{center}
\caption{{\small The marginalized likelihood as function of the model parameters $q_1$  (left panel) and $q_2$ (right panel) for Model III. The dotted curves are obtained for SNe+BAO, the dashed curves are obtained for OHD+SNe+BAO and the solid curves are obtained for OHD+SNe+BA+CMBShift.}}
\label{likelihoodplotqz3}
\end{figure}
%%%%%%%%%%%%%%%%%%%%%%

%%%%%%%%%%%%%%%%%%%%%%%%%
\begin{table}[h!]
\caption{{\small  Results of statistical analysis of Model III with different combinations of the data sets. The value of $\chi^2_{min}/d.o.f.$ and the best fit values of the parameters along with the  associated 1$\sigma$ uncertainties are presented.}}
\begin{center}
\resizebox{0.65\textwidth}{!}{  
\begin{tabular}{ c |c |c c } 
\hline
 \hline
  Data & $\chi^2_{min}/d.o.f.$  & $q_1$ & $q_2$ \\ 
 \hline
  SNe+BAO & 33.18/28 & 1.637$\pm$0.037 & 2.275$\pm$0.315\\ 

  OHD+SNe+BAO & 47.80/54 & 1.614$\pm$0.023 & 2.059$\pm$0.162\\ 

  OHD+SNe+BAO+CMBShift & 48.31/52 & 1.607$\pm$0.022 & 2.079$\pm$0.160\\ 
 \hline
\hline
\end{tabular}
}
\end{center}

\label{tablemod3}
\end{table}
%%%%%%%%%%%%%%%%%%%%%%%%%%%%%%%%%%%

%%%%%%%%%%%%%%%%%%%%%
\begin{figure}[htb]
\begin{center}
\includegraphics[angle=0, width=0.32\textwidth]{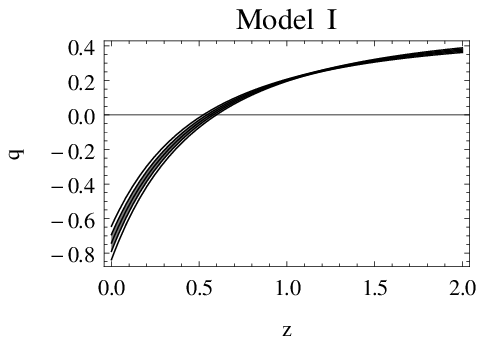}
\includegraphics[angle=0, width=0.32\textwidth]{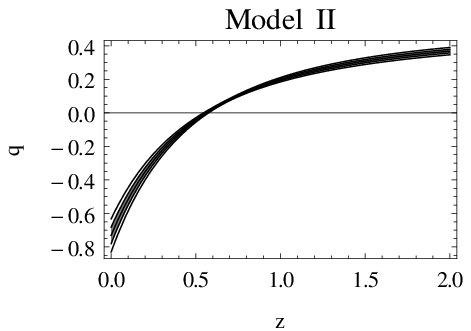}
\includegraphics[angle=0, width=0.32\textwidth]{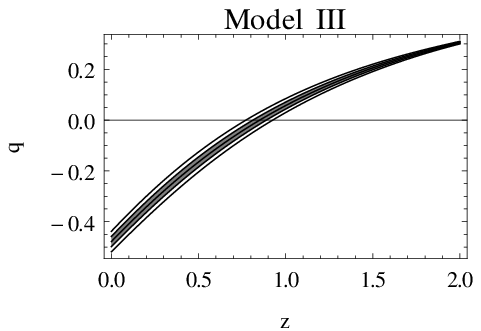}
\end{center}
\caption{{\small Plots of deceleration parameter for the models obtained from the analsis combining OHD, SNe, BAO and CMB shift parameter data. The best fit curve and the associated 1$\sigma$ and 2$\sigma$ confidence regions are presented.}}
\label{decelplot}
\end{figure}
%%%%%%%%%%%%%%%%%%%%%%

Figure \ref{contourplotqz1} shows the confidence contours on the 2D parameter space of Model I obtained from analysis with different combinations of the data sets. Figure \ref{likelihoodplotqz1} shows the plots of the marginalized likelihood as functions of the model parameters for Model I. Similarly, figure \ref{contourplotqz2} shows the confidence contours on the 2D parameter space of Model II and figure \ref{likelihoodplotqz2} shows the marginalized likelihoods of Model II. Figure \ref{contourplotqz3} and figure \ref{likelihoodplotqz3} present contour plots and likelihood plots respectively for Model III. It is apparent from the contour plots and the likelihood function plots that the addition of the CMB shift parameter data does not lead to much improvemrnt of the constraints on the model parameters. The likelihood  functions are well fitted to Gaussian distribution.  Table \ref{tablemod1} presents the results of statistical analysis of Model I. The reduced $\chi^2$ i.e. $\chi^2_{min}/d.o.f.$, where the $d.o.f.$ is the degres of freedom associated to the analysis, the best fit values of the parameters along with the associated 1$\sigma$ error bars are presented. In the similar way, table \ref{tablemod2} and table \ref{tablemod3} present the results of the statistical analysis of Model II and Model III respectively. Figure \ref{decelplot} shows the plots of deceleration parameter for the models obtained in the combined analysis with OHD, SNe, BAO and CMB shift parameter data.

\par
\begin{figure}[tb]
\begin{center}
\includegraphics[angle=0, width=0.32\textwidth]{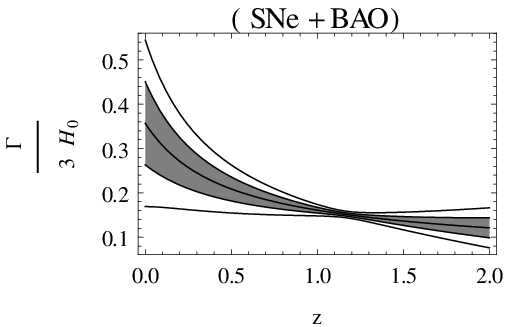}
\includegraphics[angle=0, width=0.32\textwidth]{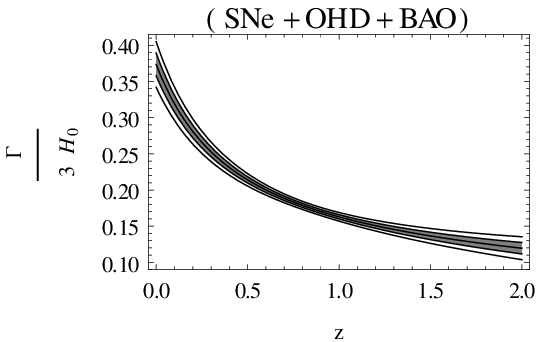}
\includegraphics[angle=0, width=0.32\textwidth]{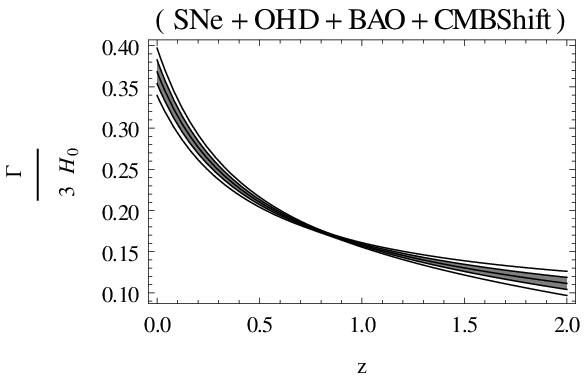}
\end{center}
\caption{{\small The plots of  interaction rate $\Gamma(z)$ scaled by $3H_0$ for Model I. Plots are obtained for three different combinations of the data sets. The left panel is obtained for SNe+BAO, the midle panel is obtained for OHD+SNe+BAO and the right panel is obtained for OHD+SNe+BAO+CMBShift. The 1$\sigma$ and 2$\sigma$ confidence regions and the corresponding best fit curves (the central dark line) are shown.}}
\label{intrateplotqz1}
\end{figure}

\begin{figure}[htb]
\begin{center}
\includegraphics[angle=0, width=0.32\textwidth]{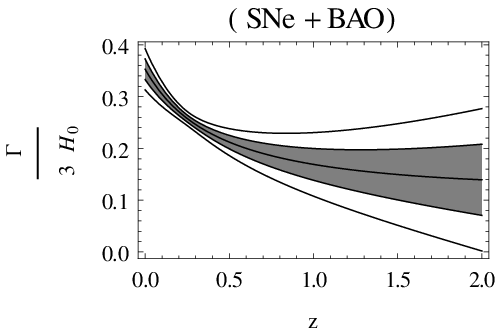}
\includegraphics[angle=0, width=0.32\textwidth]{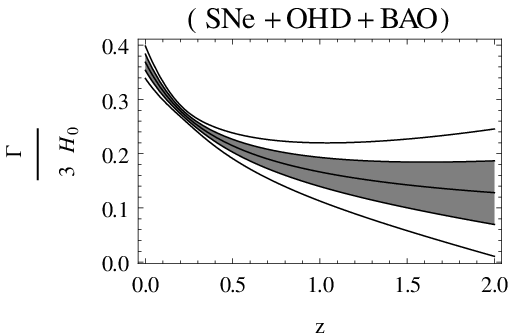}
\includegraphics[angle=0, width=0.32\textwidth]{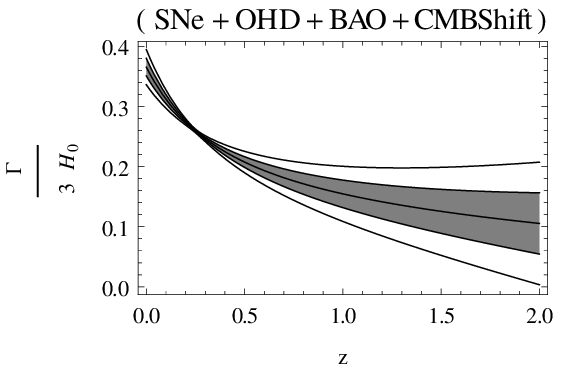}
\end{center}
\caption{{\small The plots of  interaction rate $\Gamma(z)$ scaled by $3H_0$ for Model II. Plots are obtained for three different combinations of the data sets. The left panel is obtained for SNe+BAO, the middle panel is obtained for OHD+SNe+BAO and the right panel is obtained for OHD+SNe+BAO+CMBShift. The 1$\sigma$ and 2$\sigma$ confidence regions and the corresponding best fit curves (the central dark line) are shown.}}
\label{intrateplotqz2}
\end{figure}

\begin{figure}[htb]
\begin{center}
\includegraphics[angle=0, width=0.32\textwidth]{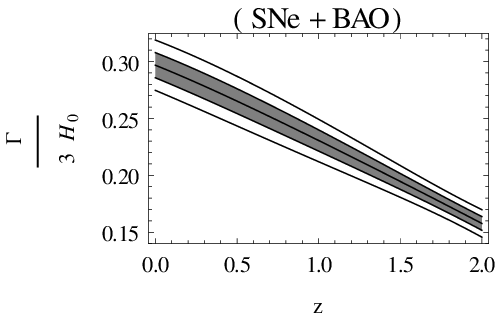}
\includegraphics[angle=0, width=0.32\textwidth]{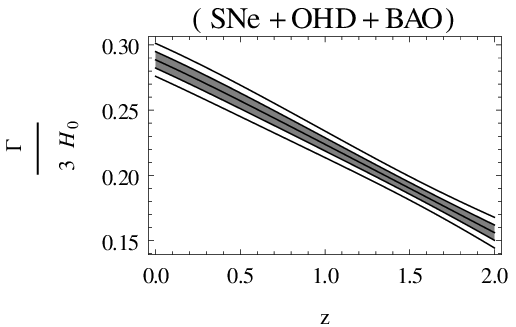}
\includegraphics[angle=0, width=0.32\textwidth]{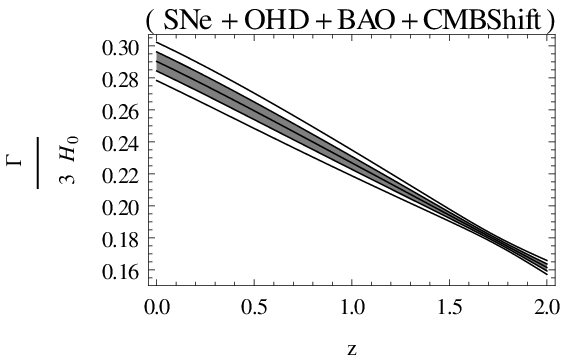}
\end{center}
\caption{{\small  The plots of  interaction rate $\Gamma(z)$ scaled by $3H_0$ for Model III. Plots are obtained for three different combinations of the data sets. The left panel is obtained for SNe+BAO, the middle panel is obtained for OHD+SNe+BAO and the right panel is obtained for OHD+SNe+BAO+CMBShift. The 1$\sigma$ and 2$\sigma$ confidence regions and the corresponding best fit curves (the central dark line) are shown.}}
\label{intrateplotqz3}
\end{figure}

 The plots of the interaction rate ($\Gamma(z)/3H_0$) (figure \ref{intrateplotqz1}, figure \ref{intrateplotqz2} and figure \ref{intrateplotqz3}) show that the interaction was low at earlier and it increases significantly at recent time. For Model I and Model III, the nature of constraint on the interaction rate, obtained in the analysis combining OHD, SNe, BAO and CMB shift data, is similar at present time and also at high redshift. But for Model II, the uncertainty increases at high redshift. 

\par
\begin{figure}[htb]
\begin{center}
\includegraphics[angle=0, width=0.32\textwidth]{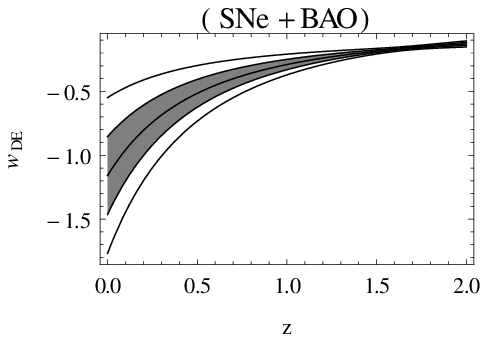}
\includegraphics[angle=0, width=0.32\textwidth]{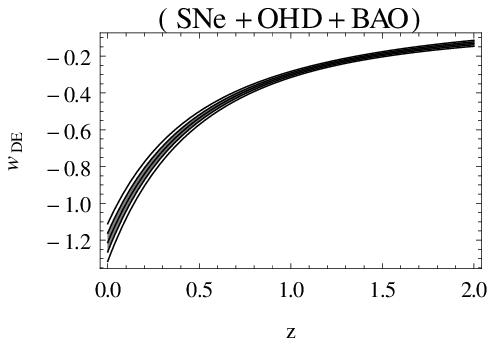}
\includegraphics[angle=0, width=0.32\textwidth]{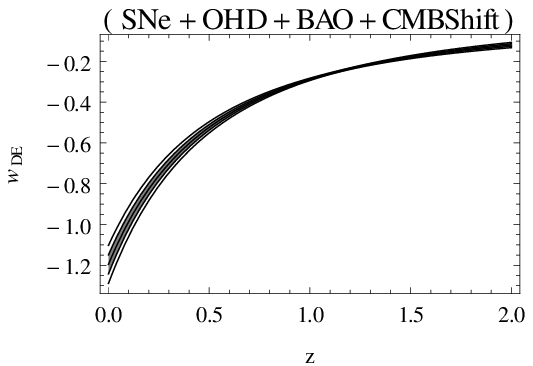}
\end{center}
\caption{{\small The plots of dark energy equation of state parameter $w_{DE}(z)$ for Model I.  The left panel is obtained for SNe+BAO, the middle panel is obtained for OHD+SNe+BAO and the right panel is obtained for OHD+SNe+BAO+CMBShift. The 1$\sigma$ and 2$\sigma$ confidence regions and the corresponding best fit curves (the central dark line) are shown.}}
\label{wDEplotqz1}
\end{figure}

\begin{figure}[htb]
\begin{center}
\includegraphics[angle=0, width=0.32\textwidth]{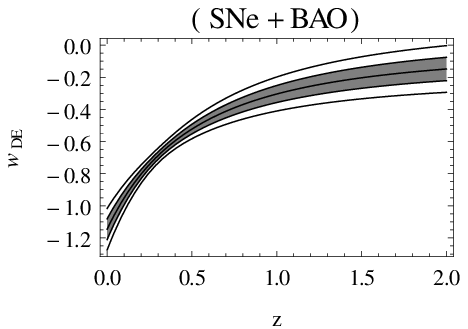}
\includegraphics[angle=0, width=0.32\textwidth]{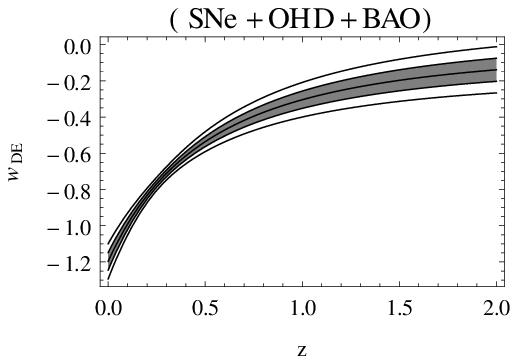}
\includegraphics[angle=0, width=0.32\textwidth]{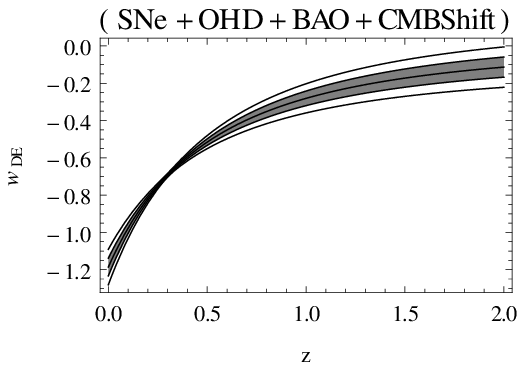}
\end{center}
\caption{{\small The plots of dark energy equation of state parameter $w_{DE}(z)$ for Model II.  The left panel is obtained for SNe+BAO, the middle panel is obtained for OHD+SNe+BAO and the right panel is obtained for OHD+SNe+BAO+CMBShift. The 1$\sigma$ and 2$\sigma$ confidence regions and the corresponding best fit curves (the central dark line) are shown.}}
\label{wDEplotqz2}
\end{figure}

\begin{figure}[htb]
\begin{center}
\includegraphics[angle=0, width=0.32\textwidth]{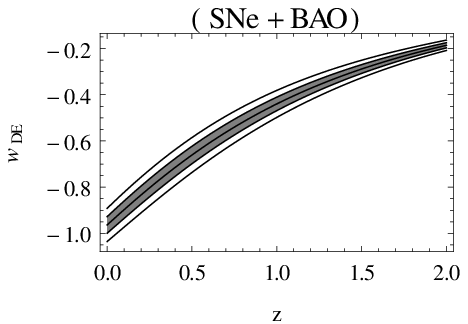}
\includegraphics[angle=0, width=0.32\textwidth]{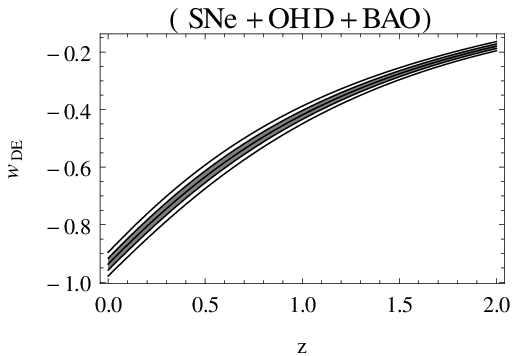}
\includegraphics[angle=0, width=0.32\textwidth]{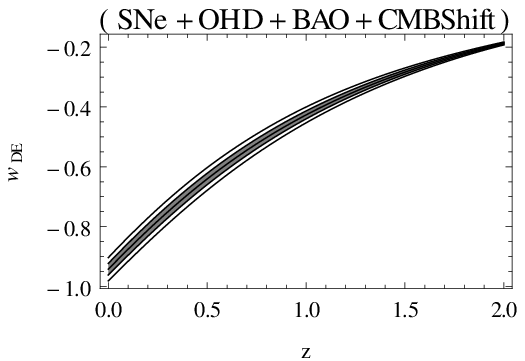}
\end{center}
\caption{{\small  The plots of dark energy equation of state parameter $w_{DE}(z)$ for Model III.  The left panel is obtained for SNe+BAO, the middle panel is obtained for OHD+SNe+BAO and the right panel is obtained for OHD+SNe+BAO+CMBShift. The 1$\sigma$ and 2$\sigma$ confidence regions and the corresponding best fit curves (the central dark line) are shown.}}
\label{wDEplotqz3}
\end{figure}

 The plots of the dark energy equation of state parameter $w_{DE}(z)$ also shows a very similar behaviour for the models, (figure \ref{wDEplotqz1}, figure \ref{wDEplotqz2} and figure \ref{wDEplotqz3}). It is imperative to note that for Model I and Model II, the dark energy equation of state parameter indicates a phantom nature at present as $w_{DE}(z=0)<-1$ at 2$\sigma$ confidence level  and for Model III, it is slightly inclined towards the non-phantom nature. At high redshift, the value of $w_{DE}(z)$ be close to zero and thus allows a matter dominated epoch in the recent past.    

\par The interaction rate $\Gamma(z)$ remains positive throughout the evolution and increases with the expansin of  the Universe. As the interaction term $Q$ is assumed to be $Q=\rho_H\Gamma$, $Q$ is also positive. This reveals that in the interaction, the energy transfers from dark energy to dark matter. It is consistent with the thermodynamic requirement of a positive $Q$ \cite{pavonwangthermo}. It is imperative to note that though the parametrization for Model III is significantly different from Model I and Model II, the basic nature of the interaction rate is same in all the case. Similar results have been obtained by Sen and Pavon \cite{senpavon} where the interaction rate of holographic dark energy has been reconstructed from a parametrization of dark energy equation of state parameter. Though tighere constraints have been achieved in the present work as it is based on larger data sets, the basic nature of the interaction rate shows no deviation from the previous findings.

\section{Bayesian Evidence and model selection}
\label{bayesian}
In the present work, three models of holographic dark energy have been discussed. It is important to look for the preferred model among these three. Two commonly used information criteria for model selection are Akaike Information Criterion (AIC) \cite{aic} and Bayesian Information Criterion (BIC) \cite{bic}. They are defined as,

%%%%%%%%%%%%%%%%%%%%%
%%%%%%%%%%%%%%%%%%%%%%
%%%%%%%%%%%%%%%%%%%%%
%%%%%%%%%%%%%%%%%%%%%%

%%%%%%%%%%%%%%%%%%%%%
%%%%%%%%%%%%%%%%%%%%%%

%%%%%%%%%%%%%%%%%%%%%
%%%%%%%%%%%%%%%%%%%%%%
%%%%%%%%%%%%%%%%%%%%%
%%%%%%%%%%%%%%%%%%%%%%

%%%%%%%%%%%%%%%%%%%%%
%%%%%%%%%%%%%%%%%%%%%%

\be
AIC=-2\log{{\mathcal{L}}_{max}}+2\kappa,
\ee
and
\be
BIC=-2\log{{\mathcal{L}}_{max}}+2\kappa\log{N},
\ee

where ${\mathcal{L}}_{max}$ is the maximum likelihood, $\kappa$ is the number of free parameter, $N$ is the number of data points used in the analysis. In the present work, all the models have two free parameters and same number of data points have been used in the analysis of the models. The difference between AIC of two models, written as $\Delta AIC$ and difference between BIC of two models, written as $\Delta BIC$, are actually the difference between the $\chi^2_{min}$ of the models. Results obtained from the analysis combining OHD, SNe, BAO and CMB shift parameter data (table \ref{tablemod1}, table \ref{tablemod2} and table \ref{tablemod3}) shows that AIC or BIC can hardly reveal any significant information regarding the selection of model among these three. So, it is useful to introduce the Bayesian evidence for model selection. The Bayesian evidence is defined as,

\be
E=\int(Prior\times Likelihood)d\theta_1d\theta_2,
\ee
where $\theta_1$ and $\theta_2$ are the parameters of the model considered. In the present analysis,  constant prior has been assumed for the parameter values for which the posterior is proportional to the likelihood. The evidence calculated for these two models are,

\be
ModelI:~~E_1=P_1\int Likelihood.dq_1dq_2=5.134\times10^{-14},
\ee
\be
ModelII:~~E_2=P_2\int Likelihood.dq_1dq_2=7.773\times10^{-14},
\ee
\be 
ModelIII:~~~E_3=P_3\int Likelihood.dq_1dq_2=21.79\times10^{-14},
\ee
where $P_1$, $P_2$ and $P_3$ are the constant prior of Model I, Model II and Model III respectively. The calculation of Bayesian evidence also does not give any significant information about the model selection as the value of $E_1$, $E_2$ and $E_3$ are not significantly different. It can only be concluded that the Model III is marginally preferred than other two models.

\section{Discussion}
\label{discussion}

The present work is an attempt to reconstruct the interaction rate for holographic dark energy. The models, discussed in this paper, are based on the parameterizations of the deceleration parameter $q(z)$. The expressions of the Hubble parameter obtained for these parameterizations of the deceleration parameter (equation (\ref{hubbleparameter1}), (\ref{hubbleparameter2}) and (\ref{hubbleparameter3})) give absolutely no indication to identify the dark matter and the dark energy components. The idea of the present work is to study the nature of interaction, mainly the interaction rate, for these three case assuming the dark energy to be holographic with Hubble horizon as the IR cut-off. As mentioned earlier, the holographic dark energy with Hubble horizon as the IR cut-off requires an interaction between dark energy and dark matter to generated the late time acceleration along with the matter dominated phase that prevailed in the past. 

\par It has also been mentioned earlier that in a spatially flat geometry, the ratio of dark matter and dark energy density in a holographic dark energy model with Hubble horizon as the IR cut-off remains constant. Thus it could be a reasonable answer to the cosmic coincidence problem. As the dark energy equation of state parameter tends to zero at high redshift, the dark energy behaved like dust matter in the past. Thus it produces the matter dominated phase in the past which is consistent with the standard models of structure formation. The interaction rate ($\Gamma$) and consequently the interaction term $Q$, where $Q=\rho_H\Gamma$, remain positive through the evolution for the reconstructed models. It indicates that in the interaction, the energy transfers from dark energy to dark matter which is consistent with the second law of thermodynamics \cite{pavonwangthermo}. Though the parametrizations are different, the basic natute of interaction rate remains same in all the cases. Similar results have also been found by Sen and Pavon \cite{senpavon} where the interaction rate has been reconstructed from parametrization of dark energy equation of state. The dark energy equation of state parameter shows a highly phantom nature at present for the Model I and Model II. For Model III, it is inclined towards the non-phanton nature.   

\par The plots of interaction rate  for these models (figure \ref{intrateplotqz1}, figure \ref{intrateplotqz2} and figure \ref{intrateplotqz3}) show that the best fit curves for Model I and Model II behave in a very similar way and for Model III it is slightly different. The nature of the associated uncertainty is different for these three models. For Model II, the uncertainty increases at high redshift. Similar behaviour can also be found in the dark energy equation of state parameter ($w_{DE}(z)$) plots of the models (figure \ref{wDEplotqz1}, figure \ref{wDEplotqz2} and figure \ref{wDEplotqz3}).

\par In the present work, three different combinations of  the data sets have been used in the analysis. The first one is  the combination of SNe and BAO, the second combination is of OHD, SNe and BAO. The CMB shift parameter data has been added to it in the third combination.  It is apparently clear that the addition of CMB shift parameter data does not lead to much improvement to the constraints on the model parameters. In case of the supernova data, the systematics have also been taken into account in the statistical analysis as the systematics might have its signature on the results. Some recent discussions on the effect of supernova systematics are discussed in \cite{supsys,supsys2,supsys3}.

\par For a comparison of models, different information criteria (namely the AIC and BIC) and the Bayesian evidence have been invoked. The Bayesian evidences calculated, are also of the same order of magnitude. It can only be concluded looking at the ratio of the Bayesian evidences of these three models that Mode III is slightly preferred than Model I and Model II, but they are comparable to each other in case of model selection.

%%%%%%%%%%%%%%%%%%%%%%%%%%%%%%%%%%%%%%%%%%%%%%%%%%%%%%%%%%%%%%%%%%%%%%%%%%%%%%
\begin{acknowledgments}
%%%%%%%%%%%%%%%%%%%%%%%%%%%%%%%%%%%%%%%%%%%%%%%%%%%%%%%%%%%%%%%%%%%%%%%%%%%%%%
The author would like to thank Professor Narayan Banerjee for guidance and valuable discussions. The author would also like to thank the anonymous referee, whose suggestions led to a substantial improvement of the paper.

%%%%%%%%%%%%%%%%%%%%%%%%%%%%%%%%%%%%%%%%%%%%%%%%%%%%%%%%%%%%%%%%%%%%%%%%%%%%%%%%
\end{acknowledgments}
%%%%%%%%%%%%%%%%%%%%%%%%%%%%%%%%%%%%%%%%%%%%%%%%%%%%%%%%%%%%%%%%%%%%%%%%%%%%%%%%

%%%%%%%%%%%%%%%%%%%%%%%%%%%%%
%\bibliographystyle{JHEP}
%\bibliography{references}

%\bibliography{references.bib}
\end{document}